\documentclass[11pt]{article}
\usepackage{amsmath,amssymb,color,graphics,epsfig,cite}
\usepackage{amsfonts}
\usepackage{xcolor}
\usepackage{multirow}

\newcommand{\be}{\begin{equation}}
\newcommand{\ee}{ \end{equation}}
\newcommand{\bea}{\begin{eqnarray}}
\newcommand{\eea}{\end{eqnarray}}

\newcommand{\hoch}[1]{$\, ^{#1}$}

\begin{document}

\begin{center}
{\Large {\bf Cosmological Time Crystals from Gauss-Bonnet Gravity in Four Dimensions}}

\vspace{20pt}

H. Khodabakhshi\hoch{1}, F. Shojai\hoch{2} and H. L\"u\hoch{1,3}

\vspace{10pt}

{\it \hoch{1}Center for Joint Quantum Studies and Department of Physics,\\ School of Science, Tianjin University,\\ Yaguan Road 135, Jinnan District, Tianjin 300350, China}

{\it \hoch{2}Department of Physics, University of Tehran, P.O.~Box 14395-547, Tehran, Iran}

{\it \hoch{3}Joint School of National University of Singapore and Tianjin University,\\
International Campus of Tianjin University, Binhai New City, Fuzhou 350207, China}

\vspace{40pt}

\underline{ABSTRACT}
\end{center}

We investigate various cosmological aspects of a 4-Dimensional Gauss-Bonnet Lagrangian, which is integrated into the Einstein Lagrangian with an arbitrary sign, using the Friedman-Lemaître-Robertson-Walker (FLRW) metric. We consider a general potential term, $V(a)$, that depends on the scale factor $a$, and we analyze several scenarios by investigating the critical points of the dynamical equations and stability conditions to understand how the universe's behavior is affected by the Gauss-Bonnet term. Our research suggests that choosing the negative sign, this integration allows for the spontaneous breaking of time reflection symmetry. This can lead to the generation of a wall-bounce universe even with a normal matter sector, marking a significant departure from traditional theories. Furthermore, we examine the possibility of a time-crystal universe, showing that under certain circumstances, the theory might give rise to cyclic universes.

\vfill{\footnotesize  h\_khodabakhshi@tju.edu.cn \ \ \ fshojai@ut.ac.ir\ \ \ mrhonglu@gmail.com \ \ \ }

\thispagestyle{empty}
\pagebreak
\tableofcontents
\addtocontents{toc}{\protect\setcounter{tocdepth}{2}}

\section{Introduction}

The concept of a time crystal, originally introduced by Frank Wilczek \cite{1,2}, involves the breaking of time-translation symmetry and the generation of periodic motion in a classical system. The emergence of time crystals challenges the conventional understanding of Hamiltonian dynamics and ground state properties. In a typical Hamiltonian system, minimizing the energy or Hamiltonian implies that the partial derivatives with respect to coordinates and momenta vanish. This condition seemingly prohibits non-trivial time or space dependence in the ground state. However, specific models with unique dynamics allow for discontinuities that violate the standard Hamiltonian equations of motion, enabling the existence of ground states exhibiting time crystal-like behavior\cite{1,2}. Theoretical investigations have explored various mechanisms through which time crystals can arise, including the introduction of a potential with a well-defined regulator that induces a non-smooth velocity reversal at the boundary\cite{3}. This unconventional behavior is not caused by a traditional potential barrier but arises from spontaneous symmetry breaking in momentum space, similar to the Higgs mechanism. Experimental validation of time crystals has confirmed their existence \cite{4}, offering profound insights into the fundamental aspects of time and symmetries \cite{5,6,7,8,9}.

In \cite{10-1,10,10-22, 10-2,10-3}, the dynamics of a cosmological spacetime were investigated by adding Gauss-Bonnet (GB) gravity in four dimensions to Einstein gravity with a plus sign. Furthermore, the time crystal behavior of an oscillating scalar field in the expanding Friedmann-Lemaître-Robertson-Walker (FLRW) universe was constructed in \cite{11,12,13}. Also, in \cite{13-1}, it is shown that cosmological models of the GB term coupled with Galileons can result in standard bouncing solutions, characterized at the bounce point by: $\mathcal{H}=0$ and $\dot{\mathcal{H}}>0$, where $\mathcal{H}$ represents the Hubble parameter \cite{13-2,odit}. It is well known that the GB Lagrangian is ghost-free, and in four dimensions, it takes the form of some special Horndeski gravity \cite{hon,15}, which also has no ghosts. Therefore, in this paper we aim to consider Einstein gravity extended with 4D-GB gravity \cite{10-22,14,16,17} with an arbitrary sign, considering the FLRW metric, where the 4D-GB term is not treated as a quantum correction to the Einstein term. Choosing the minus sign results in the lower bound for the gravitational part of the Hamiltonian, which is quite similar to the Higgs mechanism in momentum space. This mechanism gives rise to a wall-bounce universe, analogous to a ping pong ball striking a brick wall \cite{5}. In fact, when higher-order curvature invariants are included, the dynamics of classical time crystals can be emerged \cite{1,3}.

We investigates the dynamics and critical points of the FLRW metric within the 4D-GB gravity framework. By identifying critical points in the phase space and analyzing their stability, we gain valuable insights into the possible trajectories the universe can take, including scenarios involving wall-bouncing universes. Introducing a general potential term $V(a)$ in the field equations, we derive solutions for various explicit examples of $V(a)$, allowing us to investigate the behavior of the universe. Specifically, we investigate the phenomenon of wall-bounce universes, showing the distinctive characteristics of 4D-GB gravity that allows for the generation of such cosmological behavior without violating the null energy condition (NEC) for the matter sector. We also analyze the stability of the sound speed, using the definitions of effective density and pressure \cite{13-2,amani}. Our study considers minimally-coupled perfect fluid matter into the 4D-EGB Lagrangian to explore the wall-bounce univers. However, future work could extend this analysis by considering external matter, as discussed in \cite{13-1,13-2,17-2} for the standard bouncing universe, to further investigate potential instabilities. Additionally, we introduce the concept of cosmological time crystals, demonstrating the emergence of cyclic universe models under specific conditions, including negative cosmological constant.

In section 2, we calculate the GB Lagrangian in four dimensions using two approaches, taking into account the FLRW metric: one by taking the limit $D\rightarrow4$ for the GB Lagrangian in D dimensions \cite{14,15} and the other via the Kaluza-Klein reduction \cite{10-1,10,10-22,16,17}. Following this, we investigate the critial points of the Friedmann equations. Then in section 3, we obtain the properties of the resulting cosmological solutions using both the Friedmannthe equation and Hamiltonian picture. In section 4,  we present some explicit examples based on different types of the perfect fluid models. We find 4D-GB gravity without matter field contributions can generate wall-bouncing universes that preserve the NEC. In section 5, we estimate the order of magnitude for the coupling constant $\lambda$ and we examine the sound speed stability. Finally we conclude the paper in the last section.

\section{Cosmological Model in 4D-GB Gravity}
	
The action for Einstein gravity, extended by Gauss-Bonnet (GB) Gravity in D dimensions, can be written as
   \begin{equation} \label{action}
		\mathcal{A}=\int d^D x \sqrt{-g}(R+k \lambda L_{\text{\tiny{GB}}}),
     \end{equation}
where $k=\pm 1$, $L_{\text{\tiny{GB}}}=R^2-4 R_{ab}R^{ab}+R_{abcd}R^{abcd}$, and $\lambda\ge0$ represents the coupling constant. We focus our study on cosmology in the context of FLRW metric, defined as
	\begin{equation} \label{metric}
		ds^2=-dt^2+a(t)^2 (dx_1^2+dx_2^2+dx_3^2).
	 \end{equation}
Our aim is to calculate the Lagrangian of GB gravity in four dimensions by employing two distinct approaches:
	
	\begin{enumerate}
		\item Calculating the GB Lagrangian in D dimensions for the FLRW metric, then taking the limit $D\rightarrow4$ by rescaling the coupling constant of the theory according to \cite{14,15}
		\begin{equation}
			\lambda \rightarrow \frac{\lambda}{D-4}\,,\label{limit}
		 \end{equation}
		
		\item Considering the Kaluza-Klein reduction of GB gravity from general dimensions to four dimensions, which results in a theory with an additional scalar field. This approach ensures a consistent and smooth limit of the four-dimensional theory under equation \eqref{limit} \cite{10-1,10,10-22,16,17}.
	\end{enumerate}
	
Starting with the first approach, we substitute the FLRW metric into the GB Lagrangian in D dimensions, yielding
	\begin{equation}
		L_{\text{\tiny{GB}}}^{\text{\tiny{(D)}}}= -\frac{(D-1)(D-2)(D-3)(D-4)}{3}\, a^{D-5}\, \dot{a}^4,
	 \end{equation}
 with an over-dot denoting differentiation with respect to t. Taking the $D\rightarrow4$ under \eqref{limit}, the effective Lagrangian from action (\ref{action}) and in four dimensions can be obtained as
	\begin{equation} \label{lag}
		L= -6 a \dot{a}^2 -2 k \lambda \frac{\dot{a}^4}{a}-\tilde{V},
	 \end{equation}
where $\tilde{V}$ is a general potential term of the matter sector. In this paper we consider a perfect fluid matter-momentum tensor $T^a_b = \text{diag}(\rho, p, p, p)$, where
\begin{equation}
\rho = \tilde{V}(a)/(2a^{3}) \hspace{1cm}  p= -\tilde{V}'(a)/(6a^{2}),
\end{equation}
 giving rise to the equation of state $w = -a\tilde{V}'/(3\tilde{V} )$\cite{5}.
	
Another way of calculating the GB Lagrangian in four dimensions is substituting the metric (\ref{metric}) into the four-dimensional Horndeski-type scalar-tensor theory
	\begin{equation} \label{lgb}
		L_{\text{\tiny{GB}}}^{\text{\tiny{(4)}}}= \phi L_{\text{\tiny{GB}}} + 4 G^{ab} \nabla_a \phi \nabla_b \phi -4 \Box \phi (\nabla \phi)^2 +2 (\nabla \phi)^4,
	 \end{equation}
in which $G^{ab}$ is the Einstein tensor and the curvature of the internal space considered zero \cite{10-1,10,10-22,14,15,16,17}. Substituting the FLRW metric into (\ref{lgb}) leads to
	\begin{equation} \label{lgb4}
		L_{\text{\tiny{GB}}}^{\text{\tiny{(4)}}}= 2 \dot{\phi} (-2 \dot{a} + a \dot{\phi}) (2 \dot{a}^2 -2 a \dot{a} \dot{\phi} + a^2 \dot{\phi}^2),
	 \end{equation}
The $\phi$-field equation of motion reduces to
	\begin{equation} \label{eom}
		\frac{d}{dt} \big[ a^3(\dot{\phi}-\mathcal{H})^3 \big]=0,
	 \end{equation}
where $\mathcal{H}=\dot{a}/a$ is the Hubble parameter. The equation (\ref{eom}) can be integrated to give
	\begin{equation} \label{s1}
		\dot{\phi}=\mathcal{H}+\frac{C}{a},
	 \end{equation}
in which the integration constant $C$ is the scalar charge associated with the shift symmetry $\phi\rightarrow \phi +$const \cite{10-1,10}. In order to obtain the effective Lagrangain of $a$, we should not simply substitute eq. (\ref{s1}) to (\ref{lgb4}), since one is not allowed to substitute a solution with an integration constant into the Lagrangian. Instead we should substitute (\ref{s1}) into the equation of motion derived from the variation of (\ref{lgb4}) with respect to $a$. We find the resulting equation motion can be derived from the following Lagrangain
\begin{equation} \label{lgb4m}
	L_{\text{\tiny{GB}}}^{\text{\tiny{(4)}}}= \frac{-6C^4-2\dot{a}^4}{a}.
	 \end{equation}
Therefore, we have an extra term for the 4D-GB gravity  from the Hordenski type scalar-tensor theory. Equation (\ref{lag}) is now augmented to become
	\begin{equation} \label{lag1}
		L=-6 a \dot{a}^2 -2 k \lambda \frac{\dot{a}^4}{a}-V,
	 \end{equation}
in which $V=\tilde{V}+6 k \alpha/a$ and $\alpha=\lambda C^4$. The time-time and the spatial (diagonal) components of the Einstein field equations are
	\begin{equation} \label{ein}
\frac{3 \dot{a}^2}{a^2}+\frac{3 \lambda \,k\, \dot{a}^4}{a^4}-\frac{3 k \alpha}{a^4} =\rho, \hspace{2cm}		-\frac{\dot{a}^2}{a^2}-\frac{2\ddot{a}}{a} -\lambda \, k \, (\frac{4 \dot{a}^2 \ddot{a}}{a^3}-\frac{\dot{a}^4}{a^4}) - \frac{k \alpha}{a^4}= p .
	 \end{equation}
 For simplicity, we consider an effective theory where the energy density is a function of the scale factor a, and we have $\rho=\rho_m+\rho_{r}+\Lambda$, $p=p_{m}+p_{r}-\Lambda$. For example, the vacuum energy density is given by $\rho=\Lambda$, and in radiation or matter dominated universes, we have $\rho_r \sim a^{-4}$ and $ \rho_m \sim a^{-3}$ respectively.

The two equations in (\ref{ein}) can be written as Friedmann equation and acceleration equation respectively
\begin{equation} \label{f1}
3\mathcal{H}^2(1+k \lambda  \mathcal{H}^2)=\rho_{tot},
 \end{equation}
and
\begin{equation} \label{f2}
2 \dot{\mathcal{H}} (1+2 k \lambda \mathcal{H}^2)=-(p_{tot}+\rho_{tot}),
 \end{equation}
where $\rho_{tot}=\rho_m+\tilde{\rho}_{r}+\Lambda$, $p_{tot}=p_{m}+\tilde{p}_{r}-\Lambda$ and we have $\tilde{\rho}_r=\rho_r (1+ k\alpha/\alpha')$ , $\tilde{p}_{r}=\tilde{\rho}_{r}/3$, $\rho_r=3k\alpha'/a^4$, where $3k\alpha'$ is the present radiation density of the universe. The effective equation of state parameter is defined as
\begin{equation} \label{es}
	w_{eff} =\frac{p_{tot}}{\rho_{tot}}.
 \end{equation}

Using the Friedmann equation (\ref{f1}), the acceleration equation (\ref{f2}) can be rewritten as Raychaudhuri equation
\begin{equation} \label{rc}
\frac{\ddot{a}}{a}(1+2\lambda k \mathcal{H}^2)-k \lambda \mathcal{H}^4=-\frac{1}{6}(\rho_{tot}+3p_{tot}).
 \end{equation}
 Imposing different conditions on $\lambda$  and $k$, from (\ref{rc}) one can find an accelerating ($\ddot{a}>0$) or decelerating ($\ddot{a}<0$) universe under the condition $\rho_{tot}+3p_{tot}<0$ or $\rho_{tot}+3p_{tot}>0$, respectively. From equations (\ref{f1}) and  (\ref{f2}), we can derive the total energy conservation equation as
\begin{equation} \label{ec}
\dot{\rho}_{tot}+3 \mathcal{H}(\rho_{tot}+p_{tot})=0.
 \end{equation}
expressing the energy conservation throughout the universe's evolution.

For a deeper understanding of the time evolution of the system, we explore the critical points of this cosmological model. For finding these points, we conside cold dark matter model with matter ($p_m=0$) and radiation ($p_r= \rho_r/3$). The cosmological equations (\ref{f1}) and (\ref{f2}) become
\begin{equation} \label{fa1}
	3\mathcal{H}^2(1+\lambda k \mathcal{H}^2)=\rho_{m}+\tilde{\rho}_{r}+\Lambda,
 \end{equation}
\begin{equation} \label{fa2}
	- 2 \dot{\mathcal{H}} (1+2 k \lambda \mathcal{H}^2)=\frac{4\tilde{\rho}_r}{3}+\rho_m.
 \end{equation}
It is advantageous to introduce three dimensionless variables
\begin{equation} \label{var}
x=\Omega_m=\frac{\rho_m}{3\mathcal{H}^2(1+ \lambda k \mathcal{H}^2)}, \hspace{1 cm} y=\Omega_r=\frac{\tilde{\rho}_r}{3\mathcal{H}^2(1+ \lambda k \mathcal{H}^2)}, \hspace{1 cm} \Omega_{\Lambda}=\frac{\Lambda}{3\mathcal{H}^2(1+ \lambda k \mathcal{H}^2)},
 \end{equation}
where $x$, $y$ and $\Omega_{\Lambda}$ respectively represent the relative energy densities of matter, radiation, and the cosmological constant with respect to the total energy density $\rho_{tot} = \rho_m+\tilde{\rho}_r+\Lambda = 3\mathcal{H}^2(1+ \lambda k \mathcal{H}^2)$. We can now rewrite the Friedmann equation (\ref{fa1}) as
\begin{equation} \label{mf1}
	1=x+y+\Omega_\Lambda.
 \end{equation}
Since $\Omega_{\Lambda}$ can be expressed in terms of $x$ and $y$, a 2D dynamical system suffices to describe the universe's evolution. For $k=-1$ the varibles $x$ and $y$ can be negative depends on the value of $\lambda$. Hence, we make the following Poincare transformation from $(x,y)$ to $(X,Y)$
\begin{equation} \label{poan}
X=x Z, \hspace{1cm} Y=y Z, \hspace{1cm}\hbox{with}\hspace{1cm} Z=\frac{1}{\sqrt{1+x^2+y^2}},
 \end{equation}
so that $(X,Y)$ are bounded, namely
 \begin{equation} \label{poan1}
Z=\sqrt{1-X^2-Y^2} \hspace{0.5cm} \rightarrow \hspace{0.5cm} -1 \le X \le 1, \hspace{0.2cm} -1 \le Y \le 1, \hspace{0.2cm} 0 \le Z \le 1.
  \end{equation}
The equation (\ref{mf1}) can be written in terms of the new variables $X$,$Y$, $\bar{\Omega}_\Lambda=Z \Omega_\Lambda$ as
\begin{equation} \label{poan3}
Z=\bar{\Omega}_\Lambda+X+Y.
 \end{equation}
In terms of the  new variables, it follows from  the constraint $X^2+Y^2\le1$ that the physically relevant dynamics in the $(x,y)$-plane occurs within a circle of unit radius, which is the physical phase space.

Differentiating $X$ and $Y$ with respect to $\eta$(where $\eta = log a$, $d\eta = H dt$) yields the 2D dynamical system
\begin{equation} \label{xp}
X'=XZ (3X+4Y-3) +X Y^2,
 \end{equation}
\begin{equation} \label{yp}
Y'=YZ (3X-4Y-4)-Y X^2,
 \end{equation}
Here, we have assumed no interaction between (dark) matter and radiation in the conservation equation (\ref{ec}). The system has five critical points: $O = (0,0)$, $R_1 = (0,1)$, $R_2=(0,-1)$, $M_1 = (1,0)$ and $M_2=(-1,0)$. The effective equation of state parameter for our $\Lambda$CDM model is
\begin{equation} \label{eos}
	w_{eff}=1-\frac{2\dot{\mathcal{H}}}{3\mathcal{H}^2}=1+\frac{(1+k \lambda \mathcal{H}^2)(4\tilde{\rho}_r/3+\rho_m)}{(\rho_m+\tilde{\rho}_r+\Lambda)(1+2k \lambda \mathcal{H}^2)}.
 \end{equation}
The properties of each point are summarized in Table (\ref{ta1}).

\begingroup
\begin{table*}
	\caption{ Critical points of the dynamical system (\ref{xp})-(\ref{yp}) and their properties. CC refers to the Cosmplogical Constant and $\Gamma=1+k \lambda  \mathcal{H}^2$.
	}\label{ta1}
	\begin{tabular}{cccc cccc}
		\hline \hline
		Fixed Points   &  $\bar{\Omega}_{\Lambda}$    & $w_{eff}$   &     Univers Dominated By    &   Stability    \\
		\hline
		$O=(0,0)$ & $1$  & $1$ & CC  & Stable Point     \\
	    $M_1=(1,0)$ & $ \frac{\Lambda}{\rho_m}=-1$ &  1($\Gamma=0$), $\infty$($\mathcal{H}^2=0$) & Matter+Negative CC& Saddle Point     \\
	    $M_2=(-1,0)$ & $ \frac{\Lambda}{\rho_m}=1$ &  1($\Gamma=0$), $3/2$($\mathcal{H}^2=0$) & Matter+Positive CC& Unstable Point     \\
	     $R_1=(0,1)$ & $ \frac{\Lambda}{\tilde{\rho}_r}=-1$ &  1($\Gamma=0$), $\infty$($\mathcal{H}^2=0$) & Radiation+Negative CC& Unstable Point     \\
	      $R_2=(0,-1)$ & $ \frac{\Lambda}{\tilde{\rho}_r}=1$ &  1($\Gamma=0$), $5/3$($\mathcal{H}^2=0$) & Radiation+Positive CC& Unstable Point     \\
		\hline \hline
	\end{tabular}
\end{table*}
\endgroup

\begin{figure}[ht]
	\centering
	\includegraphics[width=0.4\textwidth]{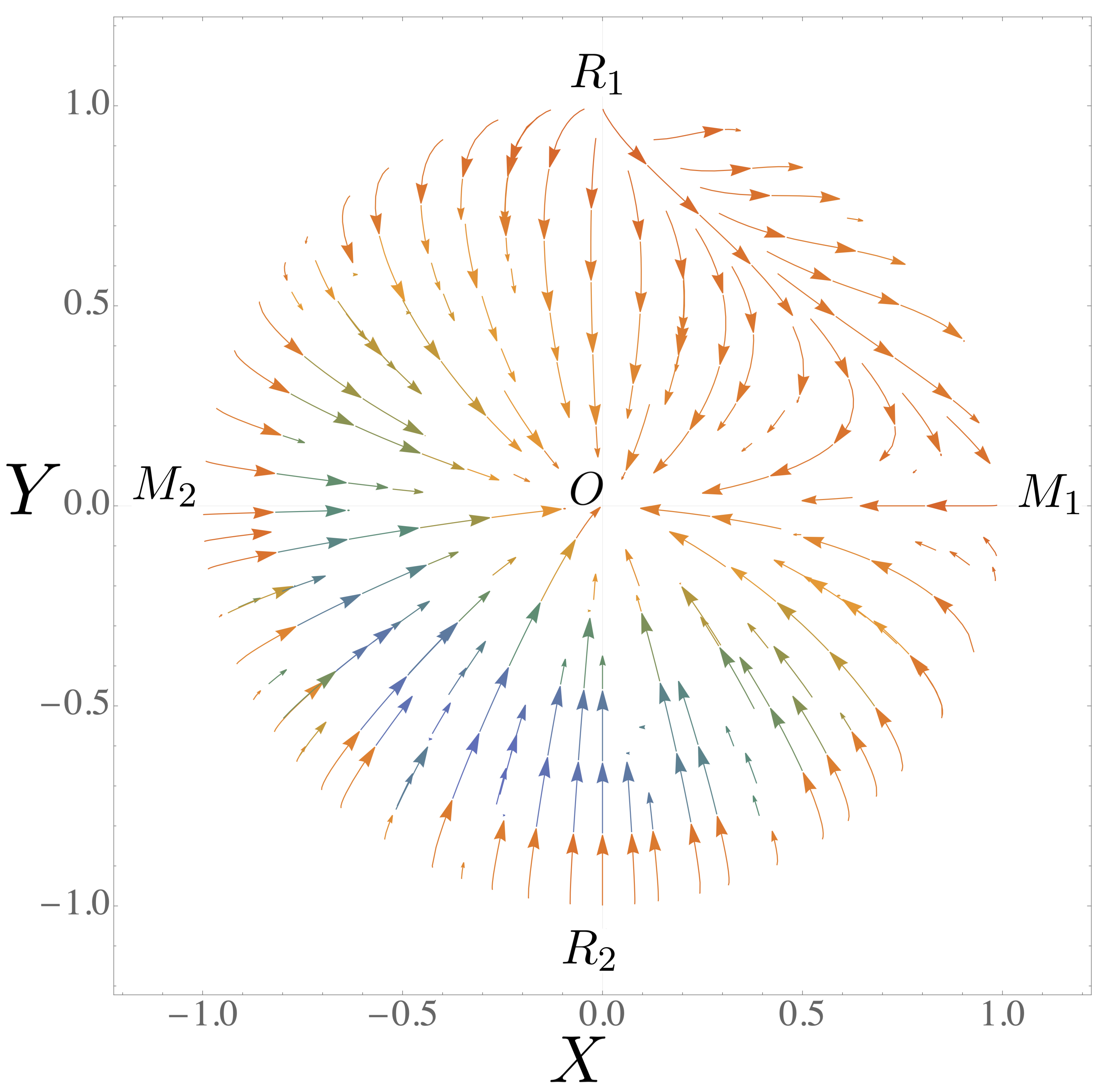}
	\caption{\small{Phase space portrait of the dynamical system (\ref{xp})-(\ref{yp}).
			The direction of the arrow at a specific point $(X,Y)$ is determined by the vestore $(X',Y')$.}}
	\label{11}
\end{figure}

The physical phase space for the system (\ref{xp})-(\ref{yp}) has been plotted in Fig. \ref{11}. The points \( R_1 \), \( R_2 \) and \( M_2\) act as past attractors, while the origin serves as the future attractor. Each solution represents a heteroclinic orbit transitioning from \( R_1 \), \( R_2 \) and \( M_2 \) (as \( \eta \rightarrow -\infty \)) to \( O \) (as \( \eta \rightarrow +\infty \)). Notable exceptions include orbits along the x-axis and the line \( Y = 1-X \), connecting \( M_1 \) to \( O \) and \( R_1 \) to \( M_1 \), respectively. Such trajectories highlight scenarios where either the cosmological constant is negative or radiation is absent.

We note that the variables  $x$ and $y$ , together with the new time variable $\eta$ require $\mathcal{H} \ne 0$. The condition $\mathcal{H} = 0$ is typically associated with an Einstein-Gauss-Bonnet static universe. Given $\mathcal{H}=0$, the Friedmann equation (\ref{fa1}) simplifies to $\rho_{m}(a_0)+\tilde{\rho}_{r}(a_0)+\Lambda=0$ and from (\ref{fa2}) we have
\begin{equation}
	\rho_m(a_0)=-\frac{4\tilde{\rho}_r(a_0)}{3},
 \end{equation}
which implies $\tilde{\rho}_r(a_0)\le0$ and $\Lambda=\tilde{\rho}_r(a_0)/3\le0$.

Morover considering the condition $1+k\lambda \mathcal{H}^2=0$ (in the case $k=-1$ we have $\mathcal{H}^2=1/\lambda$), the variables  $x$ and $y$ are not well-defined in (\ref{var}). For this case the solution of $\rho_{m}(a_0)+\tilde{\rho}_{r}(a_0)+\Lambda=0$ gives $a_0=constant$, which is a static universe and it cannot be considered as a viable model for our universe.

Another interesting point is $\mathcal{H}^2=1/2\lambda$ at $a=a_0$, where the coefficient of $\dot{\mathcal{H}}$ in equation (\ref{fa2}) is zero. In the previous analysis of the dynamical system, we cannot see any information about this point even by maximally expanding the phase space using the Poincaré transformation. The dynamics of the phase space are quite sensitive to the definition of the variables. To see this point, we define new dimensionless variables as
 \begin{equation} \label{var1}
A=\frac{\rho_m}{3\mathcal{H}^2}, \hspace{0.7cm} B=\frac{\tilde{\rho}_r}{3\mathcal{H}^2}, \hspace{0.7cm} L=\frac{\Lambda}{3\mathcal{H}^2}, \hspace{0.7cm} V=k\lambda \mathcal{H}^2.
  \end{equation}
Depending on the sign of $k$ the variables $B$ and $V$ can be either positive or negative. Now from equations (\ref{fa1}) and (\ref{fa2}), one can write:
 \begin{equation} \label{new1}
A+B+L=1+V,
 \end{equation}
and
 \begin{equation} \label{new2}
-\frac{2 \dot{\mathcal{H}}}{\mathcal{H}^2}=\frac{4A+3B}{1+2V}.
 \end{equation}
Differentiating these variables with respect to $\eta$ yields a 3D dynamical system described by
\begin{align} \label{xA}
A'&=A (\frac{3A+4B}{1+2V}) -3A,
\\\label{xB}
B'&=B (\frac{3A+4B}{1+2V}) -4B,
\\\label{xV}
V'&=-V (\frac{3A+4B}{1+2V}).
 \end{align}
The effective equation of state can be wrriten as
 \begin{equation} \label{eos1}
w_{eff}=-1+\frac{A+4B/3}{1+V}.
 \end{equation}
It should be noted that because \( \Lambda \) is a constant in equation \eqref{var1}, \( L \) can be easily expressed in terms of \( V \) as \( L = \frac{\beta}{V} \), where \( \beta = \frac{k \lambda \Lambda}{3} \). Taking this into account along with equation \eqref{new1}, it becomes evident that a 2D dynamical system suffices for understanding the key features of this cosmological model. Substituting \( L = \frac{\beta}{V} \) into equation \eqref{new1} yields

\begin{equation}
	V = \frac{1}{2} \left( (-1 + A + B) + \epsilon \sqrt{(-1 + A + B)^2 + 4\beta} \right),  \quad \epsilon = \pm 1.
	\label{ve}
 \end{equation}

\begin{figure}[ht]
	\centering
	\includegraphics[width=0.8\textwidth]{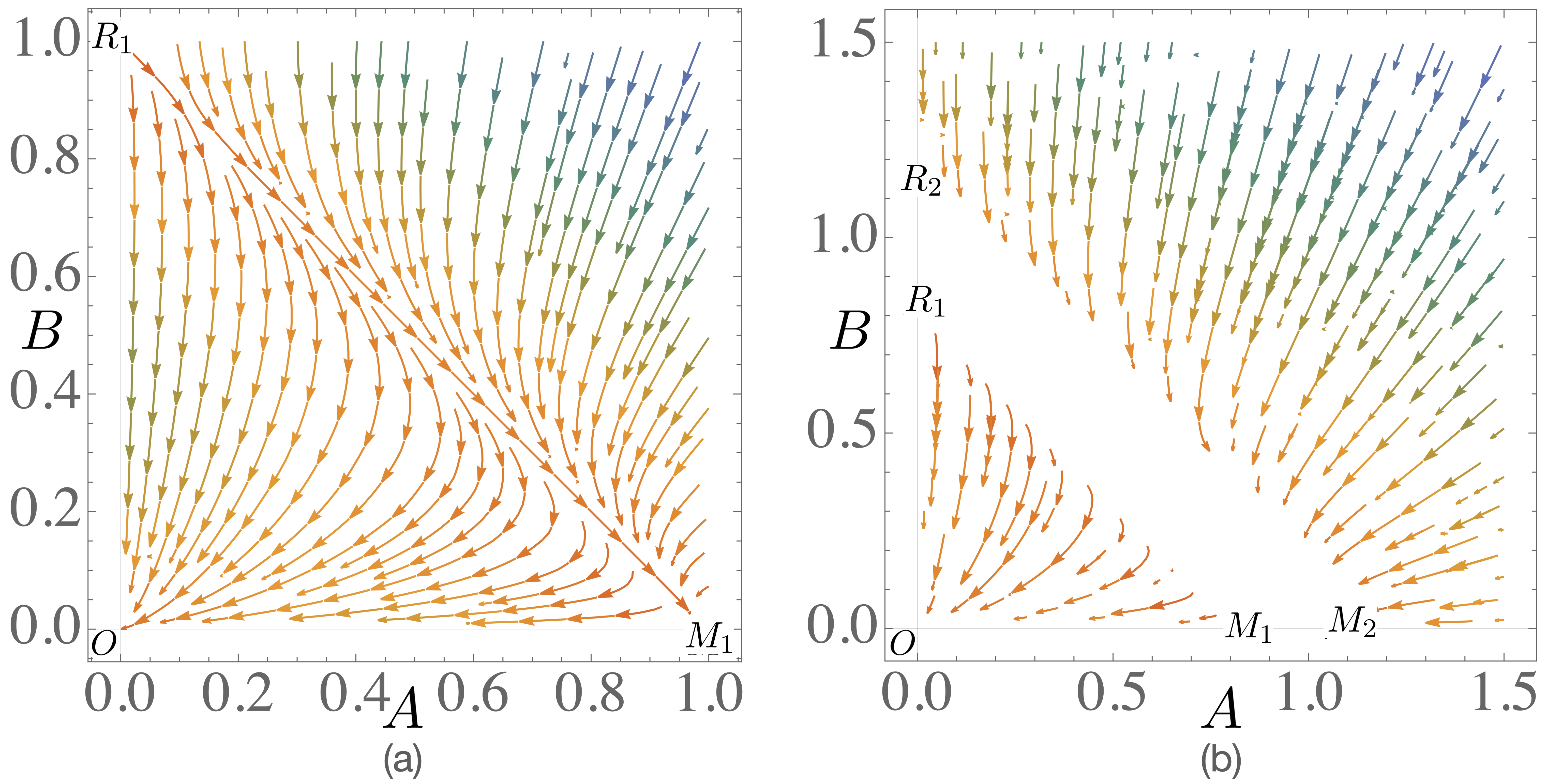}
	\caption{\small{Phase space portrait of the dynamical system \eqref{xA} to \eqref{xV}. (a) shows the case where \( \beta = 0 \), while (b) corresponds to \( \beta = -0.01 \) (\( \epsilon = 1 \)).}}
	\label{1a}
\end{figure}

\begin{figure}[ht]
	\centering
	\includegraphics[width=0.8\textwidth]{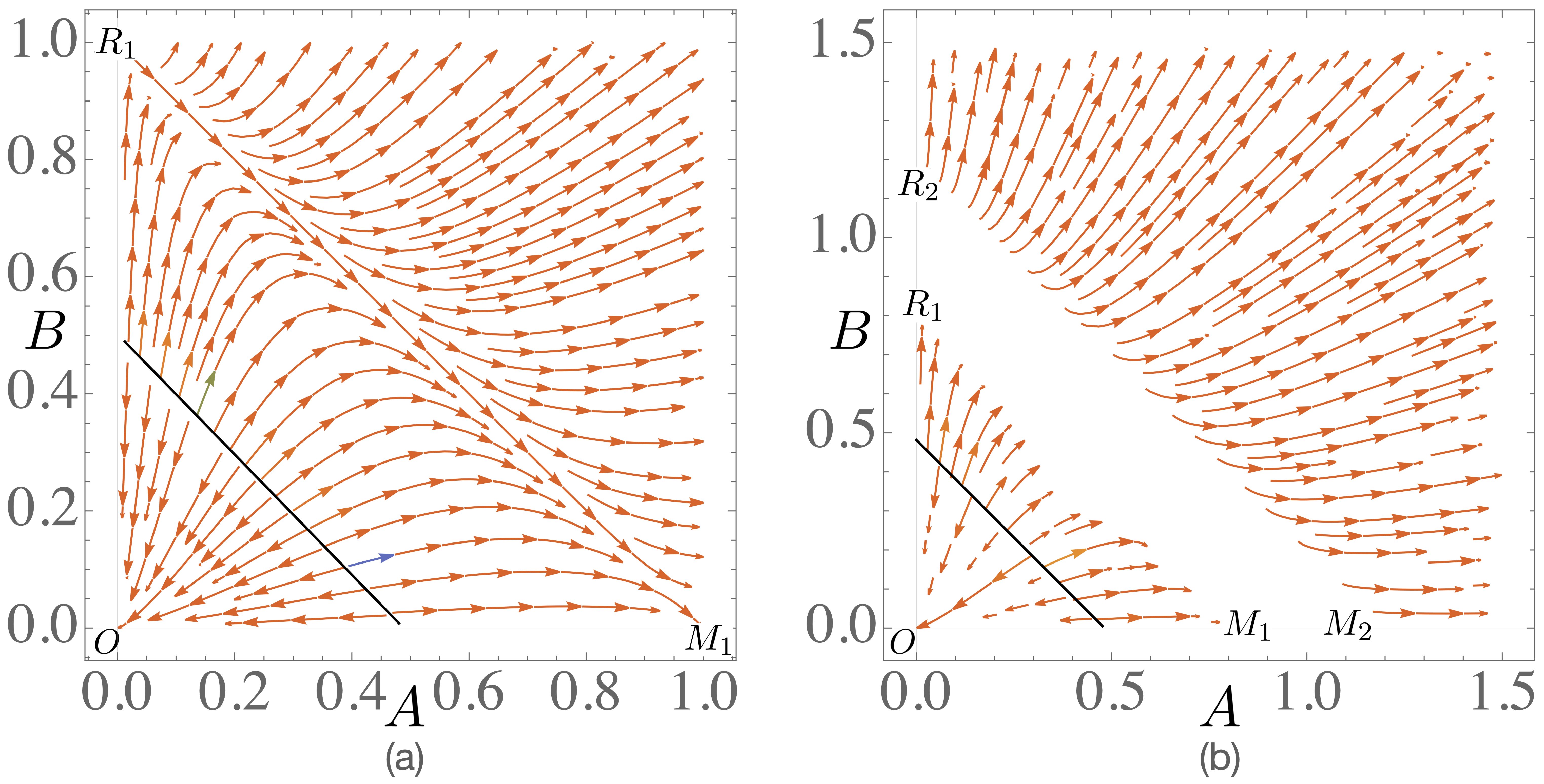}
	\caption{\small{Phase space portrait of the dynamical system \eqref{xA} to \eqref{xV}. (a) shows the case where \( \beta = 0 \), while (b) corresponds to \( \beta = -0.01 \) (\( \epsilon = -1 \)). The dashed solid lines in (a) and (b) are related to \( B = -A + \frac{1}{2} \).}}
	\label{2a}
\end{figure}

These equations provide a framework for analyzing the system's critical points, which are summarized in Table \ref{t2}. When \( \beta = 0 \) (which corresponds to \( \Lambda = 0 \) or \( \lambda = 0 \)), we have only three critical points: \( O = (0,0) \), \( M_1 = (1,0) \), and \( R_1 = (0,1) \). In this case, \( w_{\text{eff}} \) takes the values \( -1 \) at \( O \), \( 1/3 \) at \( R_1 \), and \( 0 \) at \( M_1 \). Performing linear stability analysis near these critical points, we find that the eigenvalues of the Jacobian are: \( \{-4,-3\} \) for both \( \epsilon = \pm 1 \) at \( O \), \( \{-1,3\} \) for \( z = 1 \) and \( \{-1,-3\} \) for \( \epsilon = -1 \) at \( M_1 \), \( \{4,1\} \) for \( \epsilon = 1 \) and \( \{-4,1\} \) for \( \epsilon = -1 \) at \( R_1 \). This implies that the origin is a stable point, \( R_1 \) is an unstable point for \( \epsilon = 1 \) and a saddle point for \( \epsilon = -1 \), and \( M_1 \) is a saddle point for \( \epsilon = 1 \) and a stable point for \( \epsilon = -1 \). For \( \beta = 0 \) and \( \epsilon = 1 \), the phase diagram matches that of Einstein gravity \cite{17-1} in the regions \( 0 < A < 1 \) and \( 0 < B < 1 \) (see Fig. \ref{1a}(a)). In the case where \( \beta = 0 \) and \( \epsilon = -1 \), the phase diagram shows a wall-bouncing point at \( V = -1/2 \) (\( k = -1 \) allows us to choose negative values for \( V \)) where \( B = -A + 1/2 \). Such a solution will shadow the heteroclinic orbit from the wall-bouncing point to a radiation-dominated phase, followed by a matter-dominated phase (see Fig. \ref{2a}(a)). This shows that this model can describe the evolution of the early universe originating from a wall-bouncing point.

\begin{figure}[ht]
	\centering
	\includegraphics[width=1\textwidth]{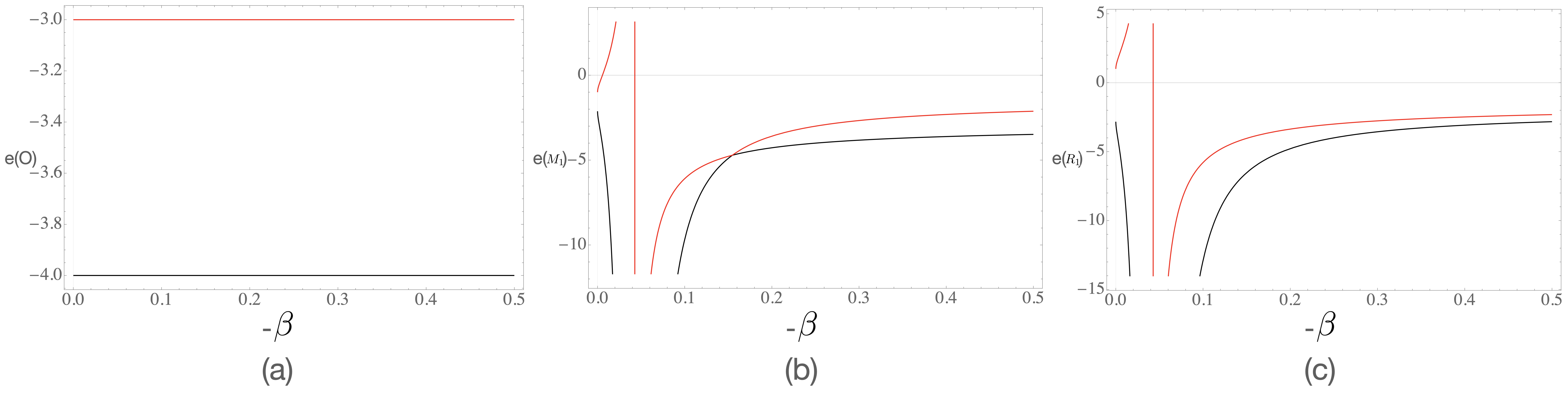}
	\caption{\small{ The black and red lines are related to the two eigenvalues \( e_{\pm} \) of the points \( O \), \( M_1 \), and \( R_1 \) as functions of  \( -\beta \) (for \( \epsilon = -1 \)). This shows how the stability of each point depends on \( \beta \).}}
	\label{3a}
\end{figure}

The degeneracy of critical points is broken by considering \( \beta \neq 0 \) and \( \beta < 0 \). In this case, we have 5 critical points shown in Table \ref{t2} (for the case \( \beta > 0 \) we only have one critical point \( O \)). From (\ref{ve}), on can find that $A$ and $B$ cannot take the values in the region where $1-2\sqrt{-\beta}<A+B<1+2\sqrt{-\beta}$. For \( k = -1 \),  \( \beta < 0 \) implies that we have a positive cosmological constant, while for \( k = 1 \), \( \Lambda \) must be negative. The phase diagrams for \( \beta = -0.01 \) have been plotted for \( \epsilon = 1 \) and \( \epsilon = -1 \) in Fig. \ref{1a}(b) and Fig. \ref{2a}(b), respectively. The stability each point depends on the value of \( \beta \). Hence, as an example, we plotted the eigenvalues of \( O \), \( M_1 \), and \( R_1 \) in terms of \( -\beta \) for the case \( \epsilon = -1 \) in Fig. \ref{3a}. This figure illustrates how, by changing $-\beta$, $M_1$ transitions from a stable point to a saddle point, while $R_1$ transitions from a saddle point to a stable point. It should be noted that for \( \beta \neq 0 \) and \( \epsilon = -1 \), the wall-bouncing universe is still possible for \( V = -1/2 \).

\begin{table*} \centering
	\caption{Critical points of the dynamical system \eqref{xA} to \eqref{xV} and their properties.}
	\label{t2}
	\begin{tabular}{cccc cccc}
		\hline \hline
		Fixed Points & \( w_{\text{eff}} \) & Universe Dominated By & Stability \\
		\hline
		\( O = (0,0) \) & \( -1 \) & Vacuum & Stable Point \\
		\( M_1 = (1-2\sqrt{-\beta},0) \) & \( -\frac{2 \epsilon \sqrt{-2\beta}}{1+(-1+\epsilon\sqrt{2})2\sqrt{-\beta}} \) & Matter & Depends on \( \beta \) \\
		\( M_2 = (1+2\sqrt{-\beta},0) \) & \( -\frac{2 \epsilon \sqrt{-2\beta}}{1+(1+\epsilon\sqrt{2})2\sqrt{-\beta}} \) & Matter & Depends on \( \beta \) \\
		\( R_1 = (0,1-2\sqrt{-\beta}) \) & \( -1+\frac{4-8\sqrt{-\beta}}{3+(-1+\epsilon\sqrt{2})6\sqrt{-\beta}} \) & Radiation & Depends on \( \beta \) \\
		\( R_2 = (0,1+2\sqrt{-\beta}) \) & \( -1+\frac{4-8\sqrt{-\beta}}{3+(-1+\epsilon\sqrt{2})6\sqrt{-\beta}} \) & Radiation & Depends on \( \beta \) \\
		\hline \hline
	\end{tabular}
\end{table*}

\section{Solutions and Hamiltonian Analysis}

For specific choices of the constant \( k \), the model can describe a universe that originates from a wall-bouncing point or one that does not. Now we will solve the Friedmann equations to further investigate the wall-bouncing point through various examples. From (\ref{f1}) one can easily derive
\begin{equation} \label{sol1}
	\dot{a}^2= -\frac{a^2}{2 k \lambda }\left(1+\epsilon\sqrt{1+ \frac{4}{3} k\lambda \rho_{\text{tot}}(a)}\right), \quad \epsilon=\pm1.
 \end{equation}
Given the vacuum solution \(\rho_{\text{tot}}=0\), selecting \(k=1\) allows only the branch \(\epsilon=-1\) of the solution, where \(\mathcal{H}=0\), which is similar to the standard Einstein case. However,  \(k=-1\) implies three vacuum solutions: \(\mathcal{H}=0\) for \(\epsilon=-1\) and \(\mathcal{H}=\pm 1/\sqrt{\lambda}\) for \(\epsilon=1\). Moreover in this case if we consider  \(\mathcal{H}^2=1/2\lambda\) at \(a=a_0\), Eq. (\ref{sol1}) can be written as
\begin{equation} \label{sol4}
	\dot{a}^2= \frac{a^2}{2 \lambda }\left(1+\epsilon\sqrt{1- \frac{\rho_{\text{tot}}(a)}{\rho_{\text{tot}}(a_0)}}\right),
 \end{equation}
where \(\rho_{\text{tot}}(a_0)=3/(4\lambda)\). Near \(a-a_0\sim 0\), this becomes
\begin{equation} \label{roott2}
	\dot{a}^2= \frac{a_0^2}{2\lambda} \left(1+\epsilon \sqrt{\frac{3}{a_0}(a-a_0) (1+w_{\text{eff}}(a_0)) }\right) + \dots.
 \end{equation}
For matter satisfying the NEC (\(w_{\text{eff}} \ge -1\)), we require \(a\ge a_0\), indicating a wall-bounce at \(a=a_0\) \cite{5} (for a study of the standard bouncing scenario in cosmology, see \cite{13-1,13-2,odit,17-2}). Additionally, at \(a=a_0\), two vacuums exist with \(\dot{a}=\pm (2\lambda)^{-1/2} a\), both being energy degenerate but breaking the time-reflection symmetry. Tunneling from one vacuum to another ensures \(\dot{a}^2\) continuity, but leads to an \(\dot{a}\) transition from negative to positive, or vice versa. This is akin to a ping pong ball hitting a brick wall.

It's worthwhile to ascertain if there is an external “brick wall” source because \(\ddot{a}\), and thus the curvature, exhibits a \(\delta\)-function singularity at the comoving time. Assuming that the turning point \(a=a_0\) is at \(t=0\), we can determine the function \(a(t)\) for small \(t\). This solution is smooth except at \(t=0\), and the additional source needed for the wall-bouncing behavior is formally provided by
\begin{equation} \label{root1}
	\rho_{\text{ext}}=0,  \quad p_{\text{ext}}= 2 \sqrt{3}\, \epsilon\, a_0^{-2} \, (2\lambda)^{1/4} \sqrt{1+w_{\text{eff}}} \sqrt{|t|} \delta(t).
 \end{equation}
Since energy-momentum conservation does not involve a time derivative of \(p\), we can effectively assume \(p_{\text{ext}}=0\). It has been shown \cite{3} that this matter source can be substituted with a well-defined regulator in classical time crystals. Hence, analyzing the energy condition of this external source becomes of paramount importance.

The physical scenario is clear. Due to spontaneous symmetry breaking, the cosmology splits into two energy-degenerate vacuums, characterized by \(\dot{a}^{\pm}\). As the universe contracts to \(a_0\), it tunnels from the \(\dot{a}^-\) vacuum to the \(\dot{a}^+\) vacuum and begins to expand, leading to a wall-bounce at \(a=a_0\). Equation $\rho_{tot}(a)-3/(4 \lambda)=0$ suggests that this wall-bounce mechanism is robust and invariably occurs for a positive energy density.

\begin{figure}[ht]
	\centering
	\includegraphics[width=1\textwidth]{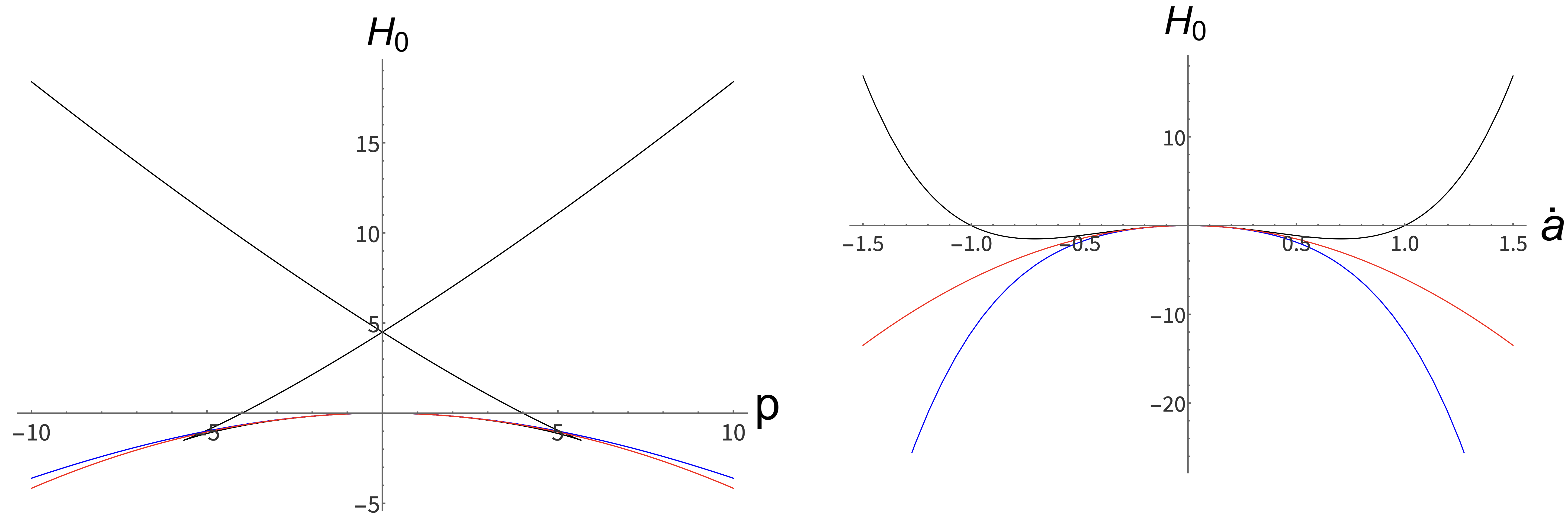}
	\caption{\small{\textbf{Black: $k=-1$, Blue: $k=1$, Red: $k=0$}. The sparrow tail shape in the left plot is characteristic when a higher-order kinetic term is included and $k=-1$. For the $k=1$ case, similar to Einstein gravity, we do not observe a sparrow tail shape; instead, it has just one maximum at $\mathcal{H}=\dot{a}=0$. The cusp singularities imply that $\dot{a}$ cannot be continuous at the tip. Instead, the system tunnels from one vacuum to the other, keeping $\dot{a}^2$ continuous. The right plot illustrates that $H_0(\dot{a})$ resembles a two-dimensional Mexican hat for $k=-1$, while this feature is absent for $k=0$ or $k=1$. It should be noted that $(0,0)$ and $(4,0)$, $(-4,0)$ points in the left plot (or equevalently $(0,0)$ and $(1,0)$, $(-1,0)$ points in the right plot) are corresponded to $\mathcal{H}=\dot{a}=0$ and  $\dot{a}=\pm (-k \lambda)^{-1/2} a$, where $H_0=0$. In both plots, we set $\lambda=1$ and $a_0=1$. For general quantities, the true vacua are at $\dot{a}=\pm (-2 k \lambda)^{-1/2} a$, corresponding to $H^{min}_0=3 a^3/2k \lambda$.}}
	\label{fig:0}
\end{figure}

The equations of motion can be derived from the effective Lagrangian (\ref{lag1}) using Hamiltonian formalism, as well. The negative kinetic energy $-\dot{y}^2/2$ proposed in \cite{1,3} is hard to justify in classical mechanics, it arises naturally in gravity. This term should not be viewed as ghost-like owing to the general diffeomorphism, which imposes the Hamiltonian constraint
\begin{equation} \label{Ham1}
	H=-6 a \dot{a}^2- 6 k \lambda \frac{\dot{a}^4}{a}+V.
 \end{equation}
This constraint is equivalent to the first equation in (\ref{ein}). In contrast, while $H$ is conserved in classical mechanics, it does not necessarily vanish. In \cite{10-1,10,10-22}, the cosmological aspects is considered for the $k=1$ case, where the Hamiltonian $H_0$ has only one maximum at $p=0=\dot{a}$, similar to the Einstein case. In this paper, we focus more on the conditions $k=-1$ and $\lambda>0$, ensuring that the gravitational part of the Hamiltonian, $H_0 = - 6 a \dot{a}^2 + 6 \lambda \dot{a}^4/a$, is bounded below. A key property is that in terms of the canonical momentum
\begin{equation} \label{mom1}
	p=\frac{\partial L}{\partial \dot{a}}=-12 a \dot{a}- 8 k \lambda \frac{\dot{a}^3}{a},
 \end{equation}
$H_0(p, a)$ can be multi-valued, and the minimum of $H_0$ does not occur when $p = 0$, but when
\begin{equation} \label{Ham01}
	\frac{\partial H_0}{\partial \dot{a}}=0 \hspace{.5cm} \Rightarrow \hspace{.5cm} \dot{a}=\pm  (-2 k \lambda)^{-1/2} a, \hspace{0.5cm} H_0= \frac{3 a^3}{2 k \lambda}.
 \end{equation}
The $p=0=\dot{a}$ point is instead a local maximum. Considering $k=-1$, this is analogous to the Higgs mechanism, but in momentum space. $H_0$ as a function of $\dot{a}$ and $p$ is depicted in Fig. \ref{fig:0} for different values of $k=0$ and $k=\pm 1$.

The evolution of $a(t)$ near $H^{min}_0$ depends on the potential $V(a)$. In order to study this behavior, we also examine $H_0$ as a function of $\dot{a}$, depicted also in Fig. \ref{fig:0}. The Hamiltonian constraint (\ref{Ham1}) at $H^{min}_0$ implies that
\begin{equation} \label{hc1}
	\bar{V}(a)\equiv V(a)-\frac{3 a ^3}{2\lambda}=0.
 \end{equation}

If it has no solution, (e.g. $V = -3a^3$, for a negative cosmological constant,) then $H^{min}_0$ can never be reached. If equation (\ref{hc1}) has a solution at $a = a_0$, the system may reach $a_0$, but cannot stay there since $\dot{a}\neq 0$. Thus the equation (\ref{Ham1}) has two positive roots, given by
\begin{equation} \label{roott1}
	\dot{a}^2=\frac{a^2}{2\lambda}\bigg( 1 +\epsilon \sqrt{1- \frac{a_0^3}{a^3}\, \frac{V(a)}{V(a_0)}}\bigg), \hspace{1.2cm} \epsilon=\pm1,
 \end{equation}
which is equevalent to the solutions in (\ref{sol4}).

The final point to discuss here is the understanding of the critical points in the Hamiltonian picture. We can express equation (\ref{Ham1}) in terms of \( A \), \( B \), \( L \), and \( V \) as
\begin{equation} \label{roottt1}
	H = H_0 \left(1 - (A+B+L-V)\right),
 \end{equation}
which implies that the Hamiltonian is zero by imposing the constrain $A+B+L-V=1$ from (\ref{new1}). In the next section, we will present some examples.

\section{Explicit Examples}

\subsubsection{Bounce universes}

\paragraph{\( \rho_{\text{tot}}=\gamma a^{b-3} \) or \( V(a)=2 \gamma a^b \):}
It should be noted that for any potential term of the form \( V(a)=2 \gamma a^b \), where $b$ is an arbitrary constant and \( \gamma \) contains \( \lambda \) and other constants, the solutions can be written as

\begin{equation} \label{rot3}
	\dot{a}^2=\frac{a^2}{2\lambda}\left( 1 +\epsilon \sqrt{1- \left(\frac{a_0}{a}\right)^{(3-b)}}\right).
 \end{equation}

Assuming \( a(0)=a_0 \), the above equation can be integrated to express \( t \) as

\begin{equation} \label{rot7}
	t=\frac{2\sqrt{\lambda}}{b-3}\left(\sqrt{2}+\text{ArcCoth}[\sqrt{2}]-\left(\frac{1}{\mathcal{B}(a)}+\text{ArcTan}[\mathcal{B}(a)] \right) \right),
 \end{equation}

where

\begin{equation} \label{rot5}
	\mathcal{B}(a)=\sqrt{\frac{1+\epsilon\sqrt{1- \left(\frac{a_0}{a}\right)^{(3-b)}}}{2}},
 \end{equation}

and the Hubble parameter is given by \( \mathcal{H}=(a(dt/da))^{-1}=(\lambda)^{-1/2} \mathcal{B}(a) \).  It shows as \( a \) nears \( a_0 \), \( t \) approaches \( \frac{2 \sqrt{\lambda} (\sqrt{2} + \text{ArcCoth}[\sqrt{2}])}{(3-b)} \) and \( \mathcal{H} \) approaches \( \frac{1}{\sqrt{2 \lambda}} \). As \( a \) approaches infinity, \( t \) goes to infinity and \( \mathcal{H} \) tend towards \( 0 \) and  \( 1/\lambda \) for \( z=-1 \) and  (\( z=1 \), \( b<3 \)) respectively.

Furthermore, using Eq. \eqref{rot3}, we can express the scale factor \( a(t) \) in terms of \( t \) near \( a-a_0 \) as

\begin{equation} \label{rot6}
	a(t)=a_0+\frac{a_0}{\sqrt{2\lambda}} |t|+ z \frac{a_0}{(2\lambda)^{3/4}}\frac{\sqrt{3-b}}{3} |t|^{3/2}+\cdots\,.
 \end{equation}

\begin{figure}
	\begin{center}
			\centering
	\includegraphics[width=0.6\textwidth]{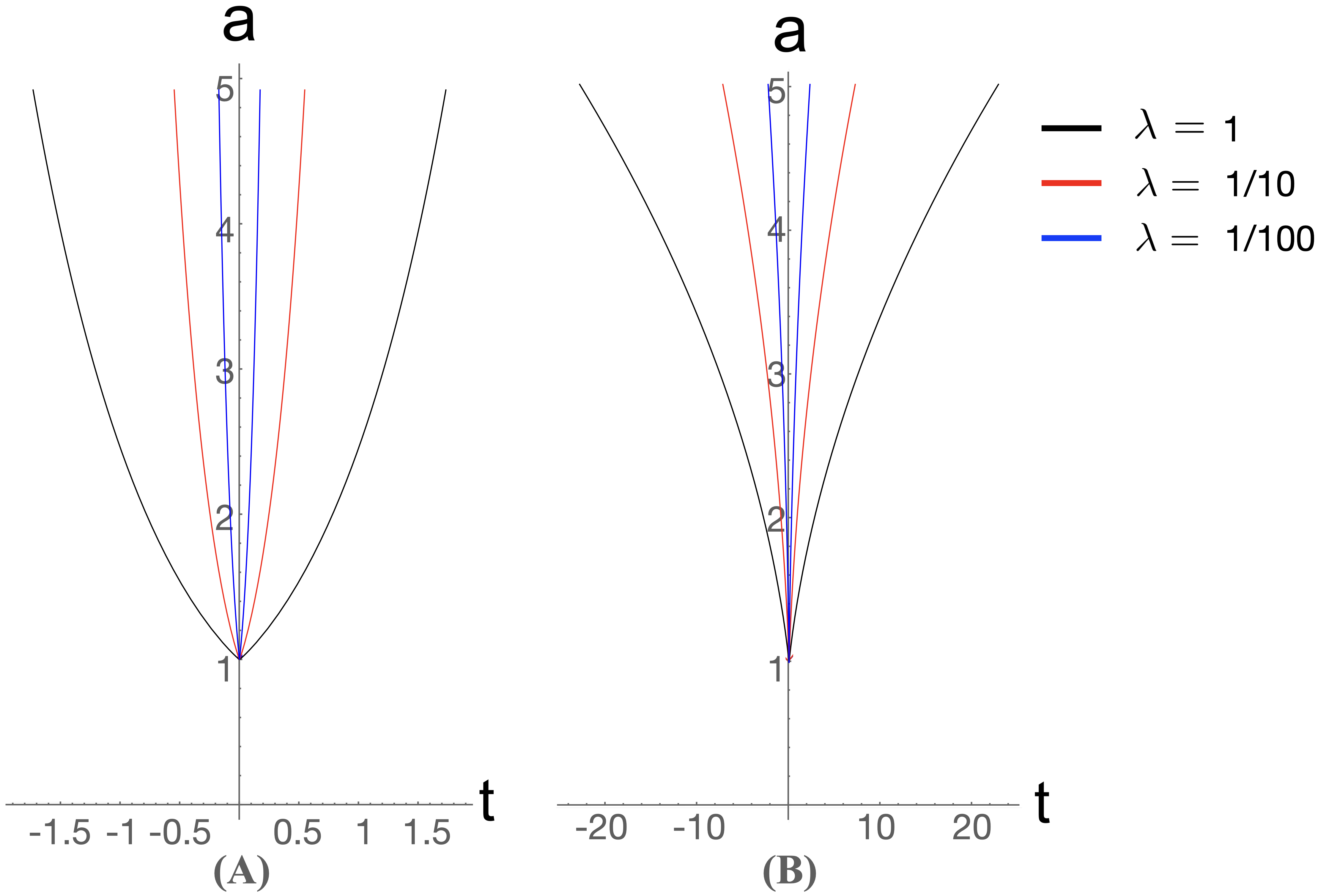}
		\caption{\small{The A ($\epsilon=1$)  and B ($\epsilon=-1$) plots  shows wall-bounce universes, where $\dot{a}$ reverses its sign at $t=0$, $a_0=1$. We have chosen different values for $\lambda$ and it shows by reducing the value of $\lambda$, the deviation of $a$ from zero will be smaller. }}
		\label{fig:1}
	\end{center}
\end{figure}

For \( b=-1 \), we are examining the scenario where  \( \rho_{\text{tot}}=\tilde{\rho_r} = 3 \alpha' (1-\alpha/\alpha') a^{-4} \). This implies that contributions from the matter field \( \rho_m \) and the cosmological constant \( \Lambda \) are zero. In the presence of a non-zero \( \lambda \), a wall-bounce is essential, occurring at \( a_0^4 = 4 \lambda \alpha' (1-\alpha/\alpha') \). This reveals that 4D-GB gravity, even without the presence of a matter field and cosmological constant, produces a wall-bounce universe within standard cosmology with a positive energy density. In Fig. \ref{fig:1}, we present the solutions for \( a(t) \) with \( a_0=1 \) for different values of \( \lambda \). Importantly, our analysis confirms the stability of these solutions against minor perturbations of the initial condition \( a(0)=1 \). It's crucial to emphasize that our proposed wall-bounce scenario diverges from those in existing studies, where \( \dot{a}=1 \) at the wall-bounce. As a result, violating the NEC is necessary within Einstein's gravity framework. In our model, \( \dot{a} \) doesn't vanish during the wall-bounce; instead, it transitions from a negative to a positive value, ensuring the NEC is upheld.

\begin{figure}
	\centering
	\includegraphics[width=0.9\textwidth]{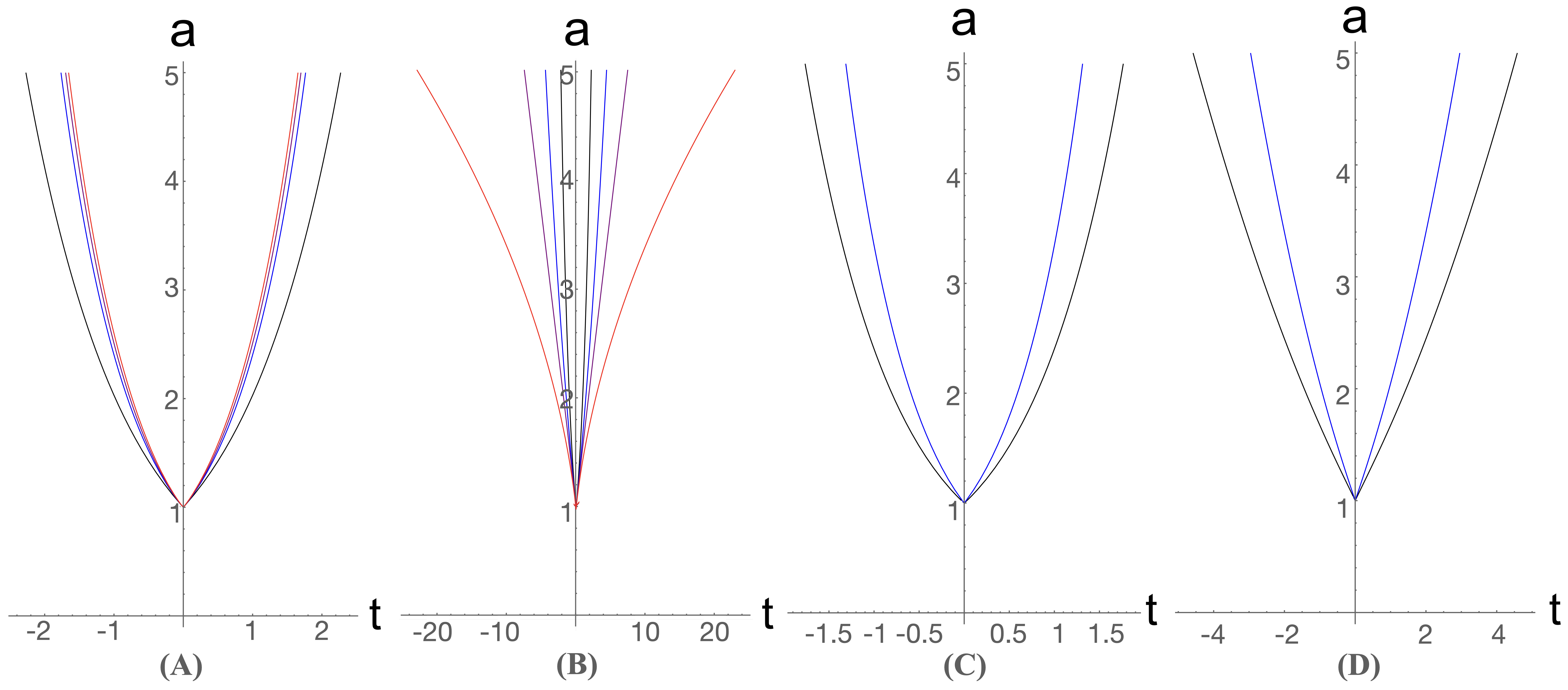}
	\caption{For \( \lambda=1 \) and \( a_0=1 \), plots A (\( \epsilon=1 \)) and B (\( \epsilon=-1 \)) represent wall-bounce universes, with \( \dot{a} \) switching sign at \( t=0 \). The colors black, blue, purple, and red correspond to \( \omega \) values of -1, -2/3, -1/3, and 1/3, respectively. Plots C (\( \epsilon=1 \)) and D (\( \epsilon=-1 \)) compare 4D-GB gravity (black line) to Einstein Cubic gravity (blue line).}
	\label{fig:2}
\end{figure}

Considering a constant \( w \), \( \rho_{\text{tot}} \) is given by \( q^2 a^{-3(w+1)} \) where \( q \) is a constant. For \( \lambda=0 \), the universe expands according to \( a=\left(\frac{3}{4} (1+\omega)^2 2q^2 t^2\right)^{1/(3(1+\omega))} \), starting with an initial spacetime singularity at \( t=0 \). For a non-zero \( \lambda \), a wall-bounce is necessary, occurring at \( a_0^{3(1+\omega)} = \frac{4 \lambda q^2}{3} \). The solution follows Eq. \eqref{sol1}, with \( b=-3w \). Similar to the \( \rho_m=\Lambda=0 \) scenario, introducing 4D-GB gravity generates a wall-bounce universe. In Fig. \ref{fig:2} A and B, we plot \( a(t) \) solutions for various values of \( \omega \), setting \( a_0=1 \) again. The solutions remain stable against slight perturbations of the initial condition \( a(0)=1 \), and similar to the radiation case, the matter system satisfies the NEC. In Fig. \ref{fig:2} C and D, 4D-GB gravity is compared with Einstein Cubic gravity for \( \omega=-2/3 \). The universe's expansion in 4D-GB gravity is slower than the Einstein Cubic case.

\subsubsection{Cosmological Time Crystals}

\begin{figure}
	\centering
	\includegraphics[width=0.8\textwidth]{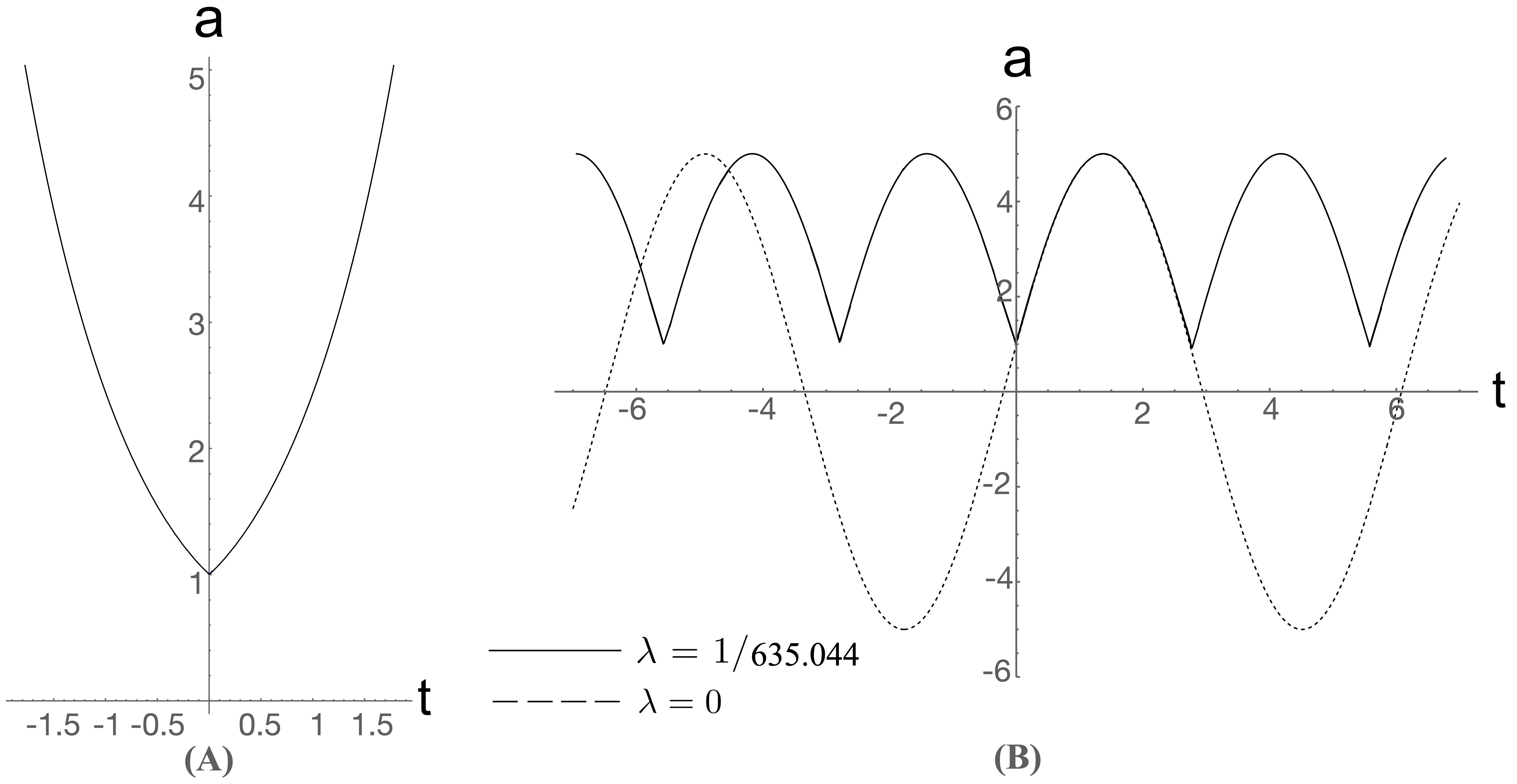}
	\caption{{\bf A}: The plot illustrates bounce universes where \( \dot{a} \) changes sign at \( t=0 \), considering \( \epsilon=1 \), \( a_0=1 \), \( \mathcal{A}=5 \), and \( \omega=-1 \). {\bf B}: The solid line on the left plot shows a cyclic universe due to the time-crystal mechanism (\( a_0=1 \), \( \mathcal{A}=5 \), \( l=1 \), \( \epsilon=-1 \), corresponding to \( \Lambda=-3 \) and \( \alpha=5 \)). The dotted line represents the exact solution for \( a \) when \( \lambda=0 \).}
	\label{fig:cos}
\end{figure}

\paragraph{\( \rho_{\text{tot}}=3 \alpha_t a^{-4}+ q^2 a^{-3 (w+1)} \) or \( V=6 \alpha_t a^{-1}+2 q^2 a^{-3w} \):}
Here, we consider the combined contributions from all terms in 4D-GB gravity, along with the potential term \( \tilde{V}=2 q^2 a^{-3\omega} \), where \( \alpha_t=\alpha'(1-\alpha/\alpha') \). Assuming the potential \( V \) becomes zero at a finite \( a=\mathcal{A}>a_0 \), and that \( \bar{V}(a)<0 \) for \( a_0<a<\mathcal{A} \), \( a(t) \) will decrease smoothly from \( a=\mathcal{A} \) to \( a=a_0 \), where it wall-bounces, resulting in a cyclic universe. The solution is given by

\begin{equation} \label{R1}
	\dot{a}^2=\frac{a^2}{2\lambda}\left( 1 +\epsilon \sqrt{1- \left(\frac{a_0}{a}\right)^{(4+3\omega)}\frac{\mathcal{A} a^{3\omega}-\mathcal{A}^{3\omega} a}{\mathcal{A} a_0^{3\omega}-\mathcal{A}^{3\omega} a_0}}\right).
 \end{equation}
In Figure \ref{fig:cos}A, we present the solution for \( z=1 \) and \( w=-1 \). The plot exhibits a wall-bouncing universe, as previously noted. Interestingly, the other graphs considering \( \omega=-2/3, -1/3 \) almost overlap with this graph as well.

For the \( \epsilon=-1 \) solution, let's examine \( V = 2\Lambda a^3 + 6 \alpha_t a^{-1} \) or \( \rho_{\text{tot}}=\Lambda+3 \alpha_t a^{-4} \), where \( \rho_m=0 \). Choosing \( \omega=-1 \) and \( q^2=\Lambda \) and assuming a negative cosmological constant \( \Lambda = -3/l^2 \), the potential becomes zero at \( \mathcal{A}=\alpha_t^{1/4} l^{1/2} \). For \( \lambda=0 \), an exact solution exists: \( a=\mathcal{A} \sin(t/l) \). However, this solution, while appearing cyclic, is not representative of the universe due to the curvature singularity at \( a=0 \). With \( \lambda \) considered, the \( \dot{a}^4 \) term doesn't influence \( a \) at \( \mathcal{A} \), where \( \dot{a}=0 \). Nevertheless, there is a turning point $a_0$:

\begin{equation} \label{R2}
	0<a_0=\left(\frac{4 \alpha_t \lambda l^2}{4\lambda + l^2}\right)^{1/4}<\mathcal{A}.
 \end{equation}

Here, the universe wall-bounces. This solution is depicted in Figure \ref{fig:cos}B. It's important to mention that for this specific model, the time crystal mechanism only works with a negative cosmological constant since \( \alpha_t \) is positive given \( \tilde{\rho_r}>0 \). A similar behavior is expected for other \( \omega \) values like \( -2/3, -1/3 \) due to the time-crystal mechanism, producing analogous cyclic universes.

As a final example, consider \( V = 2\Lambda a^3 + 6 \gamma^2 a \) or \( \rho_{\text{tot}}=\Lambda+3 \gamma^2 a^{-2} \). We have assumed \( \rho_m=\tilde{\rho}_r=0 \), and the second term arises from a sigma model \cite{18}. In this scenario, for a negative cosmological constant \( \Lambda = -3/l^2 \), the potential vanishes at \( \mathcal{A}=\gamma l \). It behaves similarly to the previous example, and there is a bounce point \( a_0 \):

\begin{equation} \label{root}
	0<a_0=\frac{2 \gamma l \sqrt{\lambda}}{\sqrt{4\lambda + l^2}}<\mathcal{A}.
 \end{equation}

Here, the universe undergoes a bounce. This solution is fitted with Figure \ref{fig:cos}B, assuming \( \lambda=1/91.711 \). Similar to the previous case, the time crystal mechanism is only possible for the negative cosmological constant, as \( \gamma^2 \) must be positive for the standard kinetic term in the sigma model \cite{18}.

\section{Valid Range of Coupling Constant and Stability Analysis}

The Friedmann equation (\ref{f1}) can be written in the following form by recaling the fundamental constants \( G \) and \( c \) as
\begin{equation} \label{eq1}
	\mathcal{H}^2 + k\lambda \mathcal{H}^4 = \frac{\Lambda c^2}{3} + \frac{8 \pi G (\rho_m + \rho_r)}{3}  + \frac{k \lambda C^4}{a^4} \, .
\end{equation}
Now, the solution (\ref{sol1}) can be expressed in terms of redshift \( z = -1 + 1/a \) and the Einstein-Hubble constant \( \mathcal{H}_{\text{\tiny{E}}} \) as follows
\begin{align}\label{ee1}
	\mathcal{H}^2(z, \lambda) =  -\frac{1 + \epsilon \sqrt{1 + 4 k \lambda \left(\mathcal{H}_{\tiny{\text{E}}}^2 (z) + k \lambda C^4 (1 + z)^4 \right)}}{2 k \lambda} , \hspace{1cm} \epsilon = \pm1,
\end{align}
where \( \mathcal{H}_{\tiny{\text{E}}}^2 \) is given by
\begin{equation}\label{ee2}
	\mathcal{H}_{\tiny{\text{E}}}^2 = \mathcal{H}_0^2 \left( \Omega_r (1 + z)^4 + \Omega_m (1 + z)^3 + \Omega_\Lambda \right),
\end{equation}
with \( \mathcal{H}_0 \) denoting the current Hubble constant, and the dark energy, matter, and radiation density parameters are defined as
\begin{equation}
	\Omega_\Lambda = \frac{c^2 \Lambda}{3 \mathcal{H}_0^2}, \hspace{1cm} \Omega_m = \frac{8 \pi G}{3 \mathcal{H}_0^2}\rho_{0m}, \hspace{1cm} \Omega_r = \frac{8 \pi G}{3 \mathcal{H}_0^2}\rho_{0r} ,
\end{equation}
where \( \rho_m = \rho_{0m} a^{-3} \) and \( \rho_r = \rho_{0r} a^{-4} \). In this section, we neglect the correction to the radiation part by considering \( C = 0 \) for simplicity, since it has no significant influence on our calculation. Moreover, for \( \epsilon = -1 \), the first term in the Taylor expansion of the solution (\ref{ee1}), where \( \lambda \mathcal{H}_{\tiny{\text{E}}} \ll 1 \), is equal to \( \mathcal{H}_{\tiny{\text{E}}} \), while the branch of the solution related to \( \epsilon = 1 \) has no Einstein limit.

Einstein gravity is well consistent with observational tests, so we require the correction \( \lambda \) term to be very small compared to Einstein gravity. Moreover, we know that in the recent universe, the cosmological constant or dark energy is dominant, so from (\ref{eq1}) we should have
\begin{equation} \label{eq2}
	\lambda \ll \frac{\Lambda c^2}{3 \mathcal{H}_0^4} \approx 4.79 \times 10^{-43} \, \, \, 1/(km^2/s/Mpc^2)^2 \approx 4.34 \times 10^{35} \, s^2\, ,
\end{equation}
where we used the values
\begin{equation}\label{value}
	\Omega_\Lambda = 0.685, \hspace{1cm} \Omega_m = 0.315, \hspace{1cm} \Omega_r = 8.24 \times 10^{-5}, \hspace{1cm} \mathcal{H}_0 = 67.4 \, km/s/Mpc,
\end{equation}
as estimated from \cite{19}. Equation (\ref{eq2}) provides the magnitude order of the upper bound of the coupling constant \( \lambda \) for both \( k = \pm1 \) cases. However, we should be careful that for high redshifts such as \( z = 10^{15} \), which is related to the age of leptons, if for example we require that
\begin{equation}\label{eq3}
	\frac{\mathcal{H}^2(z = 10^{15}, \lambda) - \mathcal{H}_{\tiny{\text{E}}}^2(z = 10^{15})}{\mathcal{H}_{\tiny{\text{E}}}^2(z = 10^{15})} \approx \pm 0.04,
\end{equation}
where plus and minus signs correspond to \( k = \mp1 \), one can easily see that \( \lambda \approx 10^{-22} \, s^2 \), which is much smaller than the upper bound. This shows we should take small values of \( \lambda \) to ensure that the 4D-GB model can be consistent with observational tests in the early universe.

Moreover, for the case \( k = -1 \), which is related to the wall-bouncing universe, if we calculate the lower bound of \( \lambda \), it will give us the redshift of the wall-bouncing point from the following relation:
\begin{equation}\label{eq4}
	1 + 4 \lambda \, k \, \mathcal{H}_{\tiny{\text{E}}}^2 (z) = 0,
\end{equation}
which implies \( \mathcal{H}^2 = 1/(2\lambda) \) from (\ref{ee1}) where we considered \( C = 0 \). To do so, the condition that the maximum density at the wall-bouncing point should be smaller than the Planck density yields \cite{20}
\begin{equation}\label{eq5}
	\lambda \gg \frac{3 \hbar G}{8 c^5} \approx 1.089 \times 10^{-38} \, s^2.
\end{equation}
Substituting this \( \lambda \) in (\ref{eq4}) gives us the magnitude order of the redshift of the wall-bouncing point \( z \approx 10^{19} \). This constraint on \( \lambda \) highlights its critical role in cosmological models, particularly in the context of the universe's expansion and its various phases. Future observational data may further refine these estimates, offering deeper insights into the nature of cosmic evolution and the fundamental constants that govern it.

In our final investigation of 4D-EGB cosmology, we focus on stability conditions by analyzing the sound speed parameter, $c_s^2$. By comparing the Friedmann equations (\ref{f1}) and (\ref{f2}) with the standard ones, i.e., $\rho_{\text{eff}}=3\mathcal{H}^2$ and $p_{\text{eff}}=-(3\mathcal{H}^2+2\dot{H})$, and recalling the fundamental constants, we express the effective energy density and pressure in terms of redshift $z$ as
\begin{align}
	8\pi G \rho_{\text{eff}} &= 3\mathcal{H}(z)^2 + 3k\lambda\mathcal{H}(z)^2 - 3k\lambda C^4(1+z)^4, \\
	8\pi G p_{\text{eff}} &= 2(1 + z)\mathcal{H}(z)\mathcal{H}'(z) - 3\mathcal{H}(z)^2 - 3k\lambda\mathcal{H}(z)^4 + 4k\lambda(1 + z)\mathcal{H}(z)^3\mathcal{H}'(z) - k\lambda C^4(1+z)^4,
\end{align}
where primes denote derivatives with respect to $z$. In this framework, the stability is examined through the sound speed squared, $c_s^2 = dp_{\text{eff}}/d\rho_{\text{eff}}$, with stability ensured for $c_s^2 > 0$ \cite{amani}. We consider adiabatic perturbations, where entropy variation $\delta S = 0$, leading to
\begin{equation}
	\delta p_{\text{eff}} = c_s^2 \delta \rho_{\text{eff}}.
\end{equation}
Differentiating the effective density and pressure with respect to $z$ yields
\begin{equation}
	c_s^2 = \frac{2k\lambda C^4(1 + z)^3 - (1+z)\mathcal{H}(1 + 2k\lambda\mathcal{H}^2)\mathcal{H}'' + \mathcal{H}'\left(2(\mathcal{H} + 2k\lambda\mathcal{H}^3) - (1 + z)(1 + 6k\lambda\mathcal{H}^2)\mathcal{H}'\right)}{6k\lambda C^4(1 + z)^3 - 3\mathcal{H}(1 + 2k\lambda\mathcal{H}^2)\mathcal{H}'},
\end{equation}
which simplifies to the Einstein case for $\lambda=0$. Using eqs. (\ref{ee1}) and (\ref{ee2}), we can express $c_s^2$ in terms of the density parameters for radiation $\Omega_r$ and matter $\Omega_m$ as
\begin{equation}
	c_s^2 = \frac{4(1+z)\Omega_r}{9\Omega_m + 12(1+z)\Omega_r}.
\end{equation}
This formulation demonstrates the expected dynamic behavior in a universe containing both radiation and matter: the sound speed decreases as the universe evolves from a radiation-dominated era ($z \rightarrow \infty$) to a matter-dominated era ($z$ decreases). Moreover, $c_s^2$ remains positive, indicating stability in the propagation of sound waves through the cosmological medium. In the early, radiation-dominated universe, $c_s^2 \approx \frac{1}{3}$, consistent with theoretical predictions for a relativistic fluid where $p=\rho c^2/3$. It would be also interesting to explore the impact of various external matter on this wall-bounce scenario (For discussions on the standard bouncing scenario, refer to \cite{13-1,13-2,18}).

\section{Conclusions}

In this paper, we have studied the cosmological aspects of 4-dimensional Einstein-Gauss-Bonnet gravity and its impact on cosmology. By analyzing the dynamical system, we identified various critical points that provide insights into the potential evolutionary paths of the universe. Our results highlight the potential of 4D-GB gravity to give rise to wall-bouncing universes, even in the absence of a matter field, since the Kaluza-Klein reduction leaves a dynamical scalar degree of freedom in four dimensions. This is notably different from most studies in the current literature, where wall-bouncing scenarios violate the NEC. In contrast, this work reveals scenarios where the bounce occurs with no violation of the NEC, as the scale factor's time derivative transitions from a negative to a positive value.

In the case where the potential term arises from the matter field, we have found that 4D-GB gravity still yields a wall-bouncing universe, once again without violating the NEC. Interestingly, the expansion rate of the universe within the 4D-GB gravity framework was found to be slower than in the Einstein cubic gravity case\cite{5}.

Furthermore, we have studied a general potential form $V(a)=\gamma a^x$. Here, our study provides concrete expressions for the Hubble parameter and the scale factor, which could be instrumental in subsequent investigations of the cosmological evolution under 4D-GB gravity.

We have shown that considering the full potential of 4D-GB gravity, along with a matter field contribution, can produce cosmological time crystal behavior. We proposed a scenario where the universe smoothly contracts from a maximum scale factor $a=\mathcal{A}>a_0$ to a minimum scale factor $a=a_0$, before wall-bouncing back. This bounce leads to a cyclic universe.

Finally, we calculated the estimated order of magnitude of \(\lambda\) to be approximately \(10^{-22} \, s^2\) for both the non wall-bouncing and wall-bouncing universe scenarios, where the deviation from Einstein gravity is about \(0.04\) in the early universe. Moreover, we obtained the lower bound of \(\lambda \approx 10^{-38} \, s^2\), which gives us the redshift of the wall-bouncing point \(z \approx 10^{19}\). We also showed the sound speed stability of the model using the effective density and pressure.The 4D-GB gravity theory can yield new and interesting results in the study of the early universe, improving our understanding of fundamental cosmological phenomena. Future work can further exploit this potential by investigating more complex potential terms and examining the observational implications of 4D-GB cosmology. Based on this paper, we have studied the observational tests of 4D-EGB gravity using Planck measurements of cosmic microwave background anisotropies and various baryonic acoustic oscillation datasets to precisely determine the valid ranges of the coupling constant $\lambda$ and the scalar charge $C$ \cite{far}.

\section*{Acknowledgement}

We are grateful to Robert Mann and Marzieh Farhang for useful discussions. This work was supported in part by NSFC (National Natural Science Foundation of China) Grants No.~11935009 and No.~12375052.

\end{document}